\documentclass[a4paper,11pt]{article}
\usepackage{pos}
\usepackage{physics}
\usepackage{gensymb}
\usepackage{xcolor}
\usepackage{enumitem}
\setlist{nolistsep}
\usepackage{setspace}
\usepackage{multicol}
\usepackage{tocbasic}
\DeclareTOCStyleEntry[beforeskip=.1em plus 1pt,pagenumberformat=\textbf]{tocline}{section}
\title{Review of Neutrino Experiments Searching for Astrophysical Neutrinos}
 \ShortTitle{Neutrino Experiments Searching for Astrophysical Neutrinos}

\author*[a]{Valentin Decoene}

\affiliation[a]{SUBATECH, IN2P3-CNRS, Nantes Universit\'e, Ecole des Mines de Nantes\\
  4 rue Alfred Kaslter, Nantes, France}

\emailAdd{valentin.decoene@subatech.in2p3.fr}

\abstract{Over the last two decades, we have intensified our search for a ghost particle, with the hope that it would provide us with information on the darkest places of our Universe. This quest has been conducted from the deep caves of the Earth, up to the upper layers of our atmosphere, and from one Pole to another.
In this review, I will summarize the odyssey of the search for astrophysical neutrinos. I will focus on the recent discoveries and technical developments that led us to the point where we stand now. I will highlight the different types of neutrino detectors, and their performances enabling the discovery of high-energy astrophysical neutrinos, and the understanding of their sources behind.
Finally, I will present some possible paths to the remaining uncharted territory of the ultra-high-energy neutrino astronomy.}

\ConferenceLogo{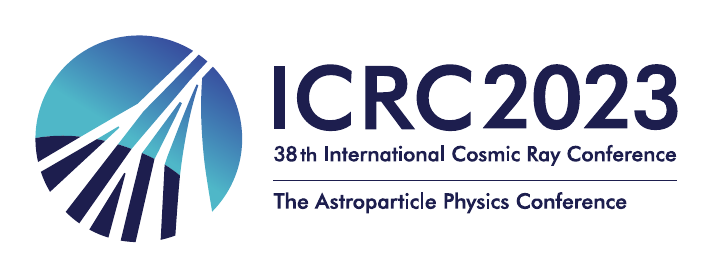}

\FullConference{%
38th International Cosmic Ray Conference (ICRC2023)\\
  26 July - 3 August, 2023\\
  Nagoya, Japan}

\tableofcontents

\begin{document}
\maketitle

\vspace{-1cm}
\section{Introduction}  \label{sec:intro}
\vspace{-0.2cm}
Neutrinos are elusive particles, that hold key answers to long-standing questions in astrophysics and fundamental physics~\cite{SnowMassHENu_2022}.
As detailed below, astrophysical neutrinos are a crucial piece of this puzzle because they cover a large energy range, way beyond what can be achieved by accelerators on Earth, and because they are expected to be produced in the most energetic sources of the Universe.

\begin{figure}[!ht]
\centering 
\includegraphics[width=0.48\columnwidth]{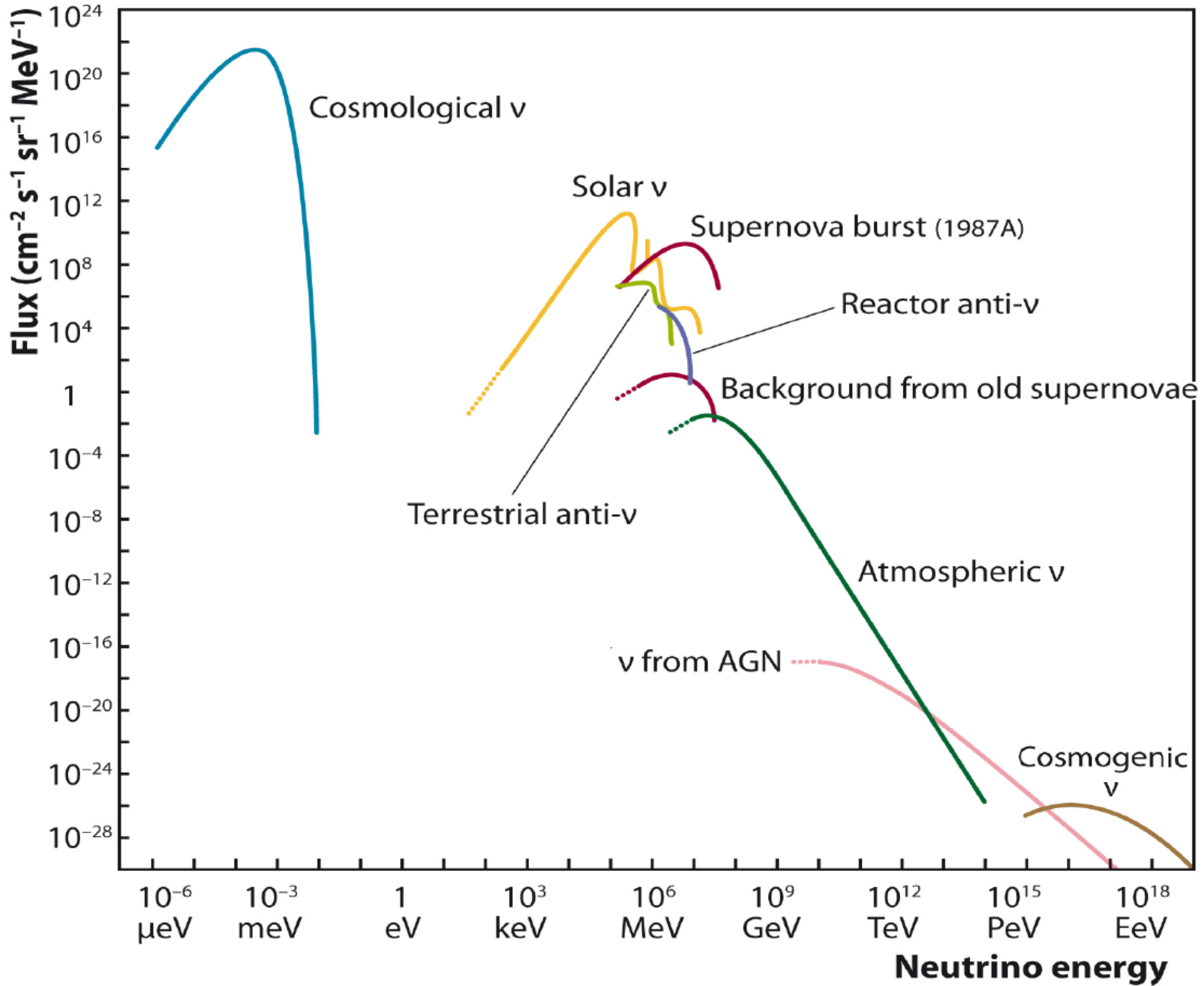}
\includegraphics[width=0.48\columnwidth]{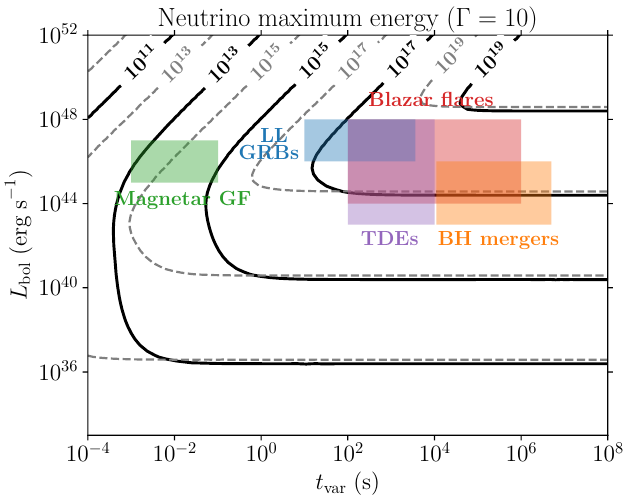}
\caption{{\it Left :} energy neutrino spectra for various types of sources (from chapter 17 of~\cite{ParticlePhysRef_2020}). {\it Right :} neutrino maximal energy, as a function of the bolometric luminosity and the time variability for different types of source. For more details, see~\cite{Guepin_2022}.\vspace{-0.2cm}}\label{fig:intro}
\end{figure}

Astrophysical neutrinos are produced in various types of environments and therefore through various production channels, leading to different expected energies. Therefore, detecting neutrinos in different energy ranges, allows for probing different types of production channels, hence, different types of environments and sources.
Typically, two main channels of neutrino production can be distinguished: beta decay reactions, and interactions of accelerated cosmic rays. These interactions, with photon and baryon backgrounds, lead to the creation of charged pions, kaons and charmed hadrons that can decay and produce subsequent neutrinos~\cite{Guepin_2022}.
This distinction, typically, draws a limit between the MeV-GeV energy range and energies above the TeV range. In reality, the two mechanisms can be interlinked and interactions of cosmic rays can lead to subsequent beta-decay reactions. However, this distinction holds in the sense that some sources cannot accelerate cosmic rays but may still produce MeV-GeV neutrinos via beta decay reactions. The left panel of Figure~\ref{fig:intro} presents an illustration of various fluxes expected from different types of neutrino sources.

For instance, in the MeV energy range, the main sources of astrophysical neutrinos are the Sun and core-collapse supernovae, both resulting from beta decay nuclear reactions. 
At higher energy, in the TeV-PeV regime, also called high-energy, neutrinos are produced through the interactions of accelerated cosmic rays, either in steady-state sources, such as active-galactic-nuclei, or transient sources. 
These sources could also produce ultra-high energy neutrinos (above PeV).
Transient sources are particularly promising candidates at the highest energies as they can inject a huge amount of energy over short time scales, thus enabling the production of astroparticles with fluxes detectable on Earth~\cite{Guepin_2022} (see the right panel of Figure~\ref{fig:intro}). 
Furthermore, in the ultra-high-energy regime, so-called cosmogenic neutrinos are also expected to be produced by ultra-high-energy cosmic rays propagating through the Universe, and interacting with photon backgrounds.
Because of their low cross sections, neutrinos are an excellent probe of the deep Universe, and can escape from opaque astrophysical sources, unlike light. Hence, they can trace processes and mechanisms hidden to standard astronomy, such as hadronic interactions and acceleration mechanisms.

From a practical point of view, their theoretically predicted energy and fluxes vary greatly from source to source. 
From the left panel of Figure~\ref{fig:intro}, it can be noticed that each source flux follows a broken power law, and the overall fluxes tend to drastically diminish with increasing energy. 
The main challenge to detect astrophysical neutrinos is to meet the detection volume required to reach those fluxes, while accommodating for the large energy ranges between the various sources. To illustrate this, let us mention that the typical required effective volume to detect Solar neutrino, in the MeV regime, is on the order of a thousand cubic meters of water, which corresponds to a kiloton of target, while in order to detect neutrinos in the TeV regime, as expected from active-galactic-nuclei, the required effective volume of water reaches a cubic kilometer, corresponding to a gigaton of water. Finally, the detection of ultra-high-energy neutrinos, as expected from cosmogenic fluxes, would require several tens of gigatons of target.
In this review, I will present how these challenges have been tackled by the community through the last few decades, and across the different energy ranges of astrophysical neutrinos.
In particular, I will summarize the detection principles that have allowed us to detect the first astrophysical neutrinos. Finally, I will present some new ideas and techniques that are studied today to pursue our exploration of the neutrino Universe towards the highest energies.

\vspace{-0.2cm}
\section{Underground detectors: a window to our local neighborhood}  \label{sec:LE}
\vspace{-0.2cm}
Neutrinos in the MeV-GeV range, have been extensively targeted by underground detectors, using water Cherenkov, liquid scintillator and radio chemical techniques (see chapter 8 of ~\cite{ParticlePhysRef_2020}). At these energies the neutrino fluxes expected from extra-terrestrial origins are the highest, still, the low interaction probability of neutrinos makes their detection challenging, regarding the backgrounds generated by natural radioactivity and cosmic-ray particles showering the Earth. Therefore, this type of detector relies on a typical volume of target on the order of hundreds to thousands of kilotons, buried in deep caves, such as old mines, in order to reduce the background generated by cosmic-ray particles.
These precursors in the quest of astrophysical neutrinos have led to breakthrough discoveries with implications covering the fields of astrophysics and particle physics. We can in particular emphasize on the detection of the Solar neutrino fluxes, the atmospheric neutrinos, the neutrino oscillation effect and the famous observation of the neutrino burst from supernova SN1987A.
The detection strategies developed by this first generation of astrophysical neutrino detectors, and their results, are presented in the following subsections.

\vspace{-0.2cm}
\subsection{Catching a ghost particle}  \label{subsec:LE_detect}
\vspace{-0.2cm}
The detection strategy to catch neutrinos in the energy regime of a few hundreds of MeV to a few GeV relies mostly on the scattering of these neutrinos with the particles of the active target used in the detector. The subsequent energy transfer from the neutrino to these particles leads to two main emission mechanisms of light: 
\begin{itemize}
\item Emission via scintillation, which happens when charged and neutral particles travel through a material, and interact via radiation interaction mechanisms. For charged particles, continuous interactions with the electrons of the material (Coulomb interactions) result in atomic excitation and ionization. 
For neutral particles, direct interactions occur, leading to proton recoil or fragment spallation, inducing a transfer of energy into the medium, and similar atomic excitation and ionization, as for charged particles (see chapter 3 of~\cite{ParticlePhysRef_2020}). 

\item Another well known emission mechanism is called Cherenkov emission and happens when charged particles move faster than the speed of light in a dielectric medium. The atoms of this medium are then polarized around the direction of propagation. In order to return to equilibrium (i.e., the ground energy state), the atoms release energy through the emission of photons. The wavefront of these light emissions move at the phase speed allowed by the medium, which is $c/n$, where $c$ is the speed of light and $n$ the refractive index of the medium. The wavefronts interfere coherently resulting in the accumulation of on-phase waves, because the disturbance, created by the moving particle, propagates at a speed greater than this phase velocity. It results in shock waves of light forming a cone oriented along the particle's propagation direction and with an aperture $\theta$ determined by $\cos{\theta}=1/\beta n$, with $\beta=v/c$ the normalized particle speed. This effect is often compared to the sonic boom produced by objects moving faster than the speed of sound~\cite{Jackson_1999}.
\end{itemize}
The light produced by either or both mechanisms is then measured via photomultiplier tubes (PMTs) from which the time and amplitude information can be used to reconstruct the neutrino properties.

However, due to the large volume of the detectors, the background noise can dominate the signal produced by the neutrinos. 
The majority of the noise stems from energetic particles produced either by the natural radioactivity of the environment, or by cosmic rays interacting within the atmosphere. The particles with sufficient energy that cross the detector can lead to scintillation or Cherenkov emissions similar to neutrino signals. Two strategies exist to control these noises: shield the detector to prevent the incoming particles to enter, and track the passing particles to discriminate them against the expected signal. The shielding method is the reason why detectors are deployed in deep underground. The  rock on-top provides a natural way to block the propagation of the majority of the particles produced by cosmic rays, at a cost of an increase of the surrounding radioactivity from the rock. The tracking method can be achieved by deploying a detector "around" the detector (often made of several sub-detectors). The set of detectors can combine different methods such as plastic scintillators, and water tanks equipped with PMTs to track the Cherenkov and scintillation light produced by the incoming particles. These surrounding detectors can also provide a subsequent shielding from the natural radioactivity.

\vspace{-0.2cm}
\subsection{Illustration with Kamiokande and Super-Kamiokande}  \label{subsec:LE_KSK}
\vspace{-0.2cm}
In the 1970s, motivated by the advent of the Grand Unification Theories, that predicted the decay of the protons and neutrons, several underground detectors were developed (see e.g., Soudan II and IMB). At that time, neutrinos were considered (mostly) as a background to understand and remove from the data.

\vspace{-0.2cm}
\paragraph{Kamiokande}
The Kamiokande II detector was installed in this context, in the old mines of the Kamioka region, and started to take data in 1986~\cite{Kamiokande_1987}.
The inner detector was composed of $3$\,kton ($2140$ tons fiduciary) of pure water, monitored by $948$ $20$-inch PMTs, arranged in a $1$\,m$^2$ step grid covering the inner part of the detector. The outer part of the detector was covered by a $123$ PMTs Cherenkov counter, used to veto against incoming particles, and absorb $\gamma$-ray emissions from the radioactivity of the surrounding rock. Additionally, a slow muon monitor was installed to veto their potential decay in the detector.
Neutrinos are detected through scattering interactions, such as $\nu + e \to \nu + e$, where the electron recoil kinematic allows for the tracking the neutrino characteristics. This channel leads to an angular reconstruction with an average error of $28\degree$. Another channel that can be used to detect neutrino consists in neutrino scattering over free protons in water $\bar{\nu}_e + p \to e^+ + n$. This channel has a cross-section of about $2$ order of magnitudes higher than the electron scattering channel, but leads to an isotropic emission of $e^+$, hence no directional information can be extracted.
Energy calibration was achieved through the measurements of muon decay into electron ($\mu \to e$), and by the use of Compton scattered electrons from $\gamma$ rays produced by a neutron source interacting with a nickel target.
A trigger was issued if at least $20$ PMTs were fired within $100$\,ns. In that case, the charge and time information from each triggered channel was recorded. Typically, the trigger efficiency was of $50\%$ for electrons with energy $5-8$\,MeV, and about $90\%$ for energies above $14$\,MeV. Consequently, the trigger rate was $0.60$\,Hz, from which $0.37$\,Hz was from cosmic muons, and the rest from radioactive contamination in the water.
The reconstruction procedure was processed in two steps: first, a reconstruction of the vertex location was achieved by triangulation from the PMTs position and times; then a fit was performed to obtain the direction of the electron.

\vspace{-0.2cm}
\paragraph{Super-Kamiokande}
The successor of Kamiokande started in 1996~\cite{Suzuki_2019}. Its concept is very similar, but bigger in size and number of sensors. It is buried at a depth of $\sim 1$\,km into the Kamioka mine in Japan. Its design is a cylinder of $42.2$\,m height and $39.6$\,m diameter. The complete detector consists of $50$\,kton of water, with $22.5$\,kton of fiducial volume. The inner part is monitored by $11000$ PMTs ($50$-cm diameter). The outer detector part, used for shielding and identification of the incoming particles, is made of a $2-3$\,m thick water layer monitored by $1885$ of the same PMTs as inside.
In addition, many improvements of the electronic were planned at the construction of the detector and also during its operation.
In particular, the new electronics allows for an improvement of detection rate of supernova bursts by a factor of $100$. Consequently, the detection efficiency for $\mu \to e$ decays reaches $100\%$ for the first microsecond, and the time accuracy is at the nanosecond level for any events.
Super-Kamiokande can detect neutrinos from $3.5$\,MeV up to $100$\,GeV, from Solar up to atmospheric neutrinos, and covers the supernova range (single and relics).
The energy resolution for solar and supernovae neutrinos amount to $14.2\%$ at $10$\,MeV, while for atmospheric single muon, it is of order $2.4\%$ and degrades with increasing energies.
The angular resolution is kinematically limited to $\sim18\degree$ for neutrinos at $10$\,MeV, but in practice due to scattering of recoiled electrons, it is closer to $\sim20\degree$. For interactions at higher energy such as $\nu_\mu + X \to \mu + Y$, the resolution is worse with $30\degree$ at $1$\,GeV, while for upward going muons it reaches down to $2\degree$.

These two detectors are historically recognized for their pioneering experimental developments and scientific results. They are to be followed by another ambitious and promising successor: Hyper-Kamiokande~\cite{DiLodovico_2017}.

\vspace{-0.2cm}
\subsection{Some loud neighbors}  \label{subsec:LE_results}
\vspace{-0.2cm}
Kamiokande and Super-Kamiokande have opened a neutrino window on our local environments. The most famous discovery is probably the neutrino burst from supernova SN1987A, followed by the solar and atmospheric neutrino oscillations. In the following, these discoveries are briefly summarized.

\vspace{-0.2cm}
\paragraph{The Galactic supernova SN1987A}
On the 23 of February 1987 at 7h35min35s ($\pm 1$\,min) a neutrino burst was observed in the Kamiokande detector (as well as IMB and Baksan). The burst of $11$ neutrinos lasted for $13$s, and the signal was consistent with energies from $7.5-36$\,MeV. The first two neutrinos pointed back at the Large Magellanic Cloud, with angles $18 \pm18\degree$ and $15 \pm 27\degree$. The association is supported by the time structure and energy distribution of the events, in addition to the correlation in direction. The events occurred $18$\,h prior to any optical detection. After correction of the detector response, it is possible to estimate the integrated neutrino flux to be about $10^{10}\,\bar{\nu}_e$\,cm$^{-2}$ for energies above $8.8$\,MeV. This can be extrapolated to the source output, and lead to a neutrino energy content of $8\times 10^{52}$\,ergs for an average neutrino energy of $\sim15$\,MeV, consistent with the theoretical expectations~\cite{Kamiokande_1987}.
The discovery led, at that time, to establish a limit on the neutrino decay lifetime on the electron neutrinos and anti-neutrinos of more than $10^{5}$\,yr$/\qty(E_\nu/m_\nu)$, where $E_\nu$ and $m_\nu$ are the neutrino energy and mass respectively. This observation shed light on the solar neutrino puzzle, still incomplete at that time.

\vspace{-0.2cm}
\paragraph{Solar neutrinos}
The Sun produces the largest neutrino flux detectable on Earth. Solar neutrinos are produced in nuclear reactions occurring in the core of the Sun (and any stars above a certain mass). The basic reaction consists in the nuclear fusion of hydrogen nuclei and is called the p-p chain $4 p \to\, ^4He + 2e^+ + 2\nu_e + 26.2 {\rm MeV}$. Other reactions may take place, depending on the composition of the star and its evolution state (see e.g., p-e-p, $^7Be$, $^8B$...)
During these reactions, most of the energy is transferred through kinematic interaction to close-by charged particles and photons. After roughly $10,000$ years, the photons escape the Sun (due to multi-scattering inside the Sun's plasma) resulting in a luminosity of $3.9\times10^{33}$\,ergs/s. 
Only $3\%$ of the energy from the nuclear reactions is carried away by neutrinos, which escapes only $2$\,s after creation. 
The discovery of Solar neutrinos was first achieved by the Homestake experiment in the 1960s, which observed only $1/3$ of the expected flux. At that time, the possible explanations were: systematic errors, issues with solar models, and oscillations (still at the hypothesis stage). In 1989 Kamiokande observed $55\%$ of the expected flux, thus confirming, at least partially, the results of Homestake Chlorine. Other experiments such as SAGE and GALLEX also confirmed the solar neutrino deficit. This was called the solar neutrino puzzle. In 2001, two studies were published by the Super-Kamiokande collaboration on the solar neutrino flux and spectrum, however the results were not significant enough to conclude on the oscillation of neutrinos from the Sun. Shortly after, the SNO experiment (a $1$\,kton water Cherenkov experiment in Canada), sensitive only to electronic neutrinos, combined its results with Super-Kamiokande, reaching a significance of $3.3\,\sigma$ on the solar neutrino oscillation hypothesis.
Kamiokande and Super-Kamiokande brought critical experimental data towards the understanding of the physics at place in our Sun.
Nowadays, there is still remaining statistical uncertainty, such as the day/night effect on the neutrino flux (oscillation in matter effects), that remains to be completed.

\vspace{-0.2cm}
\paragraph{Atmospheric neutrinos} 
In the 1980s, many experiments started to study the neutrino oscillation effects. Most of them used neutrinos produced by accelerators and reactors, but could not find any evidence of neutrino oscillations. This can be explained by the fact that the neutrino energy range was about $1$\,GeV ($1$\,MeV) and the flight length about $1$\,km ($<100$\,m) for accelerator experiments (reactor experiments). Since the oscillation probability is given by $P\qty(\nu_\mu \to \nu_\mu) \propto \sin^2\qty(1.27 \delta m^2\, L/E_\nu)$, the probed parameter space for masses is $\delta m^2\gtrsim 0.1-1$\,eV$^2$, which is too large. In order to measure smaller values of mass, larger oscillation lengths are required. Neutrinos generated in the atmosphere by cosmic-ray interactions, via meson decay $\pi/K \to \mu + \nu_{\mu} \to e + 2 \nu_\mu + \nu_e$ (taking into account pion/kaon charges and neutrino/anti-neutrino states), can travel up to the Earth diameter ($\sim12700$\,km)~\cite{Kajita_2010}. The first atmospheric neutrino experiments started in the 1960s, with in particular two experiments located in deep mines in South Africa and India. 
In 1978, the first results were published, which showed no evidence of neutrino oscillation, partly due to the large systematic effects on the data. During the meantime, in the 1970s, the Grand Unified Theories predicted the decay of nucleons, and several large volume experiments started to try to detect it: for instance, the Kamiokande and IMB experiments, for which atmospheric neutrinos were a background that needed to be correctly understood in order to search for nucleon decays. 
In 1986, the first evidence of lack of muon decays in the expected signal was brought to the community, and latter confirmed by other experiments. 
It was only in 1998, with the successor of Kamiokande, Super-Kamiokande, that the official announcement of the discovery of neutrino oscillations in the atmospheric neutrino flux was made at the 18th International Conference of Neutrino Physics and Astrophysics. 
The precise measurements of the ratio of muon and electronic neutrino fluxes, and the zenith distributions (which is independent of the flux models), showed a deficit of upward-going neutrinos, only explained by neutrino oscillation, with a significance of $6\,\sigma$~\cite{Suzuki_2019}. These results with deep implications in particle physics were possible thanks to atmospheric neutrinos, produced by cosmic particles. Nowadays, the detailed study of these oscillations, and the questions of mass hierarchy between the different flavors of neutrinos remain to be answered. New experiments, such as JUNO~\cite{JUNO_2015} or KM3NeT (see section~\ref{sec:HE}), intend to shed light on them.

In fine, the underground detectors are limited in size, with a maximal cross-section of the order of $\sim 1000$\,m$^2$, for practical reason. Thus, they cannot reach the neutrino fluxes from cosmic accelerators in the TeV-PeV range.

\vspace{-0.2cm}
\section{Large scale detectors: a glimpse at the high-energy neutrino sky} \label{sec:HE}
\vspace{-0.2cm}
The fluxes of astrophysical neutrino become drastically low when the energy increases, because the sources are located farther away from us. Indeed, at lower energies the neutrino fluxes are dominated by nearby sources such as the Sun and the atmospheric neutrinos. In addition, Galactic supernovae can be considered as relatively nearby sources. However, the sources producing neutrinos at higher energy are dispersed over large distances across the Universe, as for instance active-galactic-nuclei. Therefore, the fluxes are drastically diluted across the Universe.
Consequently, the required volume of active target in order to statistically detect neutrinos in the energy range of TeV to PeV, becomes so large that it is no longer possible to scale up the Kamiokande or Super-Kamiokande techniques (see section~\ref{subsec:LE_KSK}). Therefore, large scale detectors are required, which make use of large natural active targets such as sea, ice, or lakes. These allow for a scale up of the underground detection technique, based on water Cherenkov light, in order to reach the low fluxes expected in the neutrino high-energy range.

\vspace{-0.2cm}
\subsection{Getting bigger}
\vspace{-0.2cm}
The first principles of such detection strategies were proposed by M. Markov in 1960~\cite{Markov_1961}.
The idea relies on the charged current and neutral current interactions of a neutrino with a nucleon of the target: $\nu_l + X \to l + Y$, and $\nu_l + X \to \nu_l + Y$, with $l=e, \mu, \tau$. In the case of a charged current interaction, the produced lepton $l$ emits a light track from Cherenkov emission, while in the case of a neutral current interaction, the produced nucleus $Y$ can decay and also leave a Cherenkov light imprint (due to the charged particles in the cascade). These Cherenkov lights can be measured thanks to PMTs, similarly as described in section~\ref{subsec:LE_detect}.

The charged particles produced during the neutrino interaction, can travel up to the detector and then either cross it, leaving a light track, or decay within it (or nearby), also leaving a light signature. Therefore, the effective volume is much larger than the detector volume itself.
In particular, muon tracks with upward going directions, can efficiently be identified as neutrino signatures, since no other particle can cross the Earth at these energies.
This detection strategy is well suited for high-energy neutrinos, in particular since (see chapter 17 of~\cite{ParticlePhysRef_2020}):
\begin{itemize}[noitemsep]
\item both the neutrino cross-section and the muon range increase with energy, leading to larger effective volumes at higher energies.
\item the mean angular deviation between the neutrino direction and the muon direction goes as $E^{-0.5}$, hence a better tracking of the sources and a better reconstruction at higher energies (with a better discrimination between up-going/down-going events),
\item above TeV energies, the Cherenkov light yield increases significantly, allowing for a better reconstruction of the muon energy, as good as $\sigma\qty(\log{\qty(E_\mu)})\sim0.3$, hence a better estimate of the neutrino energy (through the unfolding of the lepton spectra into a neutrino spectra).
\end{itemize}
Above $\sim 100$\,TeV, the Earth becomes opaque to neutrinos, leading to neutrino induced muons arriving preferentially near the horizon. At energies in the EeV regime, the opacity of the Earth makes neutrino induced muons to arrive from the horizon, reducing greatly the field of view. 
However, muons from astrophysical neutrino of energies in the PeV-EeV energy range, deposit a large amount of energy, which can be used as a handle to distinguish them from atmospheric muons, on a statistical basis. In addition, down-going muon tracks or cascades starting within the detector volume are necessarily produced by neutrinos.

In order to reach the required effective volumes to detect the high-energy neutrino fluxes, large volumes of natural targets, such as glaciers, lakes and seas, are used. 
The general design consists of PMTs housed in pressure glass-transparent spheres, called optical module (OM), and spread over a large volume of ice/water along strings deployed in ice or anchored at the bottom of the sea or lake. The sphere spacing is typically between $10-25$\,m, and the strings are spaced between $60-200$\,m. Compared to underground detectors, this is larger by many factors. This technique allows for the coverage of a large volume of target, but makes the detector insensitive to events with energy below $\sim10$\,GeV, due to light absorption, except if a denser sub-array is present.

For such kinds of detector, there are three main sources of background for the identification of astrophysical neutrinos: 
\begin{itemize}[noitemsep]
\item atmospheric muons: down-going muons produced by atmospheric cosmic-ray interactions,
\item random background: PMT dark counts, $^{40}K$ decay in seawater, and bioluminescence in water,
\item atmospheric neutrinos: produced in cosmic-ray interactions in the atmosphere.
\end{itemize}
The background generated by atmospheric muons and traveling downward to the detectors can be shielded by deploying the detector in deep locations below the water (sea and lakes) or ice surface. This shielding method follows the underground detector shielding technique, with however a lesser efficiency due to the lower density of the water and ice compared to rock. Therefore, most of the detectors are deployed at least below a depth of $\sim1$\,km to suppress the majority of the background from atmospheric muons. 
In addition, the reconstruction of the arrival direction can help discriminate against the atmospheric muons.
Similarly to rock, water contains radioactive isotopes such as the potassium $^{40}K$ responsible for a constant background. In addition, transient luminescence phenomenon can happen in water, mostly from biological sources. These random backgrounds can be mitigated by using veto based on conditions of local coincidences between PMTs.
Finally, atmospheric neutrinos can be separated from astrophysical neutrinos only on a statistical basis. The neutrino fluxes from astrophysical sources are expected to be harder than the atmospheric neutrino fluxes, leading to a higher signal to background discrimination as the energy increases. Down-going atmospheric neutrinos, interacting within the detector volume, can be rejected by looking for accompanying muons (muon bundle) produced in the same shower.
It has to be noted that atmospheric neutrinos provide a convenient way to calibrate the detector since their fluxes are related to cosmic rays, which have been extensively characterized in this energy range. Furthermore, at lower energies, they are used to study the physics of neutrinos, such as oscillations or mass hierarchy, similarly to underground detectors.

\vspace{-0.2cm}
\subsection{High-energy neutrino detectors}
\vspace{-0.2cm}
During the last decades, intense efforts have been pursued in order to achieve the detection of high-energy neutrinos, and open a window on the high-energy Universe. The amazing results that we have witnessed these last two decades have been possible thanks to the developments of several pioneering experiments started more than $40$ years ago.
The origins, of the current large scale detectors of high-energy neutrinos, root back to a concept developed in the 1960s, and called DUMAND for Deep Underwater Muon And Neutrino Detector. It was located close to Hawai'i, about $30$\,km away from Big Island, at a depth of $4.8$\,km. In 1993, $24$ OMs were deployed but failed due to water leakage. This drawback and financial issues caused the ending of the project in 1996 (see chapter 17 of~\cite{ParticlePhysRef_2020}).

From this starting point, various concepts emerged and deployment was investigated in various locations, such as ice caps, deep lakes and seas.

\vspace{-0.2cm}
\paragraph{Ice experiments}
The concept of detectors deployed in deep ice was investigated on the American side with {\bf AMANDA} ({\bf Antarctic Muon And Neutrino Detection Array}). The ice layer, of $3$\,km thickness at the South Pole, is used as a target and a detection medium. It is located a few hundred meters away from the Amundsen-Scott station, and is still within the actual IceCube array (see the next paragraph).
A first shallow test in 1993-1994, at a depth of $800-1000$\,m showed that at these depths the effective length is only $40-80$\,cm, because of remnant bubbles, making the reconstruction of muon tracks impossible.
Subsequent ice measurements, at different depths and locations, found that the bubbles would disappear below some depth. Therefore, a second test array was deployed at $1500-2000$\,m, where the scattering length is about $20$\,m, much worse than in water but enough to perform track reconstructions. The array was then gradually increased until its final size in 2000, with $19$ strings and $677$ OMs. If the scattering length is rather small, on the other hand, the absorption length is larger than in water, allowing for a better photon collection from Cherenkov light, hence a better sensitivity. The energy threshold was of $\sim 50$\,GeV.
However, the resolution on the angular reconstruction for muon tracks was about $2-2.5\degree$ only, because of the strong ice scattering, which blurred the photon information from the Cherenkov light cones. It was even worse for the reconstruction of cascades, with about $25\degree$ on the angular resolution. 

The last analysis from AMANDA used $6595$ neutrinos, collected during the period of $2000-2006$. It was, finally, switched off in 2009.
During the operating time of AMANDA, it was found that the ice properties would improve greatly below a major layer of dust, located at around $2000-2100$\,m depth. This motivated the deployment of a larger scale detector following the design of AMANDA.

The {\bf IceCube} collaboration successfully built the first gigaton neutrino detector, thanks to the experience and development of its precursor AMANDA. It is located at the South Pole, and was completed in December 2010. It is composed of $5160$ Digital Optical Modules (DOMs) along $86$ strings between $1450-2450$\,m depth in Antarctica's ice. In addition, $320$ DOMs are placed at the surface in the IceTop array just above each string.
Each string is composed of $60$ DOMs, with a $10$-inch PMT, connected by pairs in order to perform fast local coincidence triggering.
AMANDA was integrated as a low energy sub-detector, and then replaced by DeepCore: a high density $6$ strings sub-array in deep ice, at the center of the IceCube array.
The energy threshold is about $\sim300$\,GeV for IceCube, and $\sim 10$\,GeV for Deep Core~\cite{Guepin_2022}.
PMT pulses are sent to the surface, but only local coincidence ends in full waveform, in order to reduce the data throughput from noise hits. The data rate is about $100$\,GB/day, written on tapes. Online processing and analysis, performed on local computer farms, extracts the interesting events, such as: up-going muons, cascades, high-energy events, coincidences between IceTop and IceCube, follow-up events, etc. The refined data amount for about $20$\,GB/day and are sent via satellites communications to the outer world.
Calibration can be achieved thanks to LEDs placed on the flasher board. The DOMs cannot be retrieved from the ice since they are frozen in it, however the main digital electronic board is made of a Field Programmable Gate Array (FPGA), where new functionalities can be uploaded. This enables a continuous update of the triggering and recording capacities of the sensors. Each DOM has a local clock pulse, synchronized every few seconds to a central GPS clock, allowing for a $\sim 2$\,ns time resolution.
The muon track angular resolution is about $1\degree$ at $1$\,TeV, and below $0.5\degree$ above $10$\,TeV. In particular, very deep ice of better quality improves greatly the reconstruction performances.
On the other hand, cascades have a limited angular resolution of $\sim 10\degree$~\cite{Guepin_2022}, mostly due to scattering from the ice.
IceCube is the only detector of its kind to benefit from the co-location of a surface detector, permanently operating. The IceTop array is made of tanks filled with ice, equipped of 2 DOMs, and placed at the top of each string. The array detects air-showers, from which the coincident detection of muons by IceCube allows for the calibration of the angular absolute pointing and angular resolution of IceCube. 
IceTop can measure the energy spectrum of air-showers, up to $10^{18}$\,eV, and
the mass range of the primary particles can be estimated thanks to the combination of both detectors (sensitive both to electron and muon components of the air-shower).
IceCube can detect supernovae neutrino bursts, thanks to its low dark count (static noise). It can measure the faint increase of count rate resulting from millions of MeV neutrino interactions, as expected from the wake of neutrinos produced by a supernova. The detector records counting rate every millisecond, therefore, even a supernova from the Large Magellanic Cloud (e.g., SN1987A) would be detected with a significance of $5\,\sigma$, early enough to trigger the Super Novae Early Warning System (SNEWS).
Thanks to criteria on the configurations and vetoing conditions, High Energy Starting Events (HESE) can be safely discriminated against atmospheric neutrinos and muons. This sub-category of events alone has evidenced over $6$ years of data the existence of an astrophysical neutrino flux with more than $7\,\sigma$ of significance.

IceCube is taking data in its final configuration since January 2011, with a duty cycle larger than $99\%$, detecting every year about $10^5$ neutrino events, from which $99.9\%$ are atmospheric. The failure rate is of 1 DOM every year over the 5160 DOMs in total.

\vspace{-0.2cm}
\paragraph{Lake experiments}
The deployment of neutrino detectors in deep lakes was initiated by the {\bf Baikal Neutrino Telescope NT200}. It was a prototype array of $192$ OMs installed in the southern part of Lake Baikal, and completed in 1998. A few extensions and upgrades followed the initial prototype, until the Baikal collaboration initiated a stepwise installation of a cubic kilometer scale detector array in Lake Baikal. 

The project is called the {\bf Baikal Giant Volume Detector} ({\bf Baikal-GVD}). It is based on a modular structure of several clusters. The first cluster was deployed in 2016 at $4$\,km offshore. The detector volume increases at a rate of $1-2$ clusters per season, and since 2023, 12 clusters have been deployed.
It is currently the largest water Cherenkov detector in operation in the Northern Hemisphere. The first phase of GVD consists of $8$ clusters totaling $2304$ OMs for a detector volume of about $0.3-0.4$\,km$^3$, followed by a second phase that will consist in a total volume of $1-2$\,km$^3$.

Each cluster, is a fully functional sub-detector, working both in standalone and full array modes. They are made of $8$ strings arranged in a circle with one string in the center. The OMs are placed on vertical strings anchored at the bottom of the lake with $36$ OMs per string, hence $288$ per cluster.
The OMs are composed of a $10$-inch PMT, and spaced by $15$\,m, with the lowest OM at a depth of $1275$\,m ($100$\,m above the bottom of the lake), and the highest at $750$\,m below the surface.
The clusters are separated by $300$\,m.
A pair of neighboring OMs in coincidence can issue a local trigger, which is sent as a request to their section. The request from the three sections arrive at the Control Module (CoM) of the string, which transfers to the Cluster DAQ (data acquisition system), and produces a global trigger.
Calibrations are performed thanks to LEDs and laser stimulation. The LEDs are located in each OM, and perform amplitude and time calibrations of the OMs and between different sections. High power lasers are located within each cluster in order to both calibrate the whole cluster and the adjacent clusters. Finally, the OMs are positioned thanks to an acoustic system within each cluster and with 4 acoustic modems per string, allowing for a positioning accuracy of about $2$\,cm.
The energy threshold is $\sim 100$\,GeV, the mean angular resolution for tracks is $<1\degree$ and $4.5\degree$ for cascades, both above $10$\,TeV~\cite{Guepin_2022}.

A total of 10 cascades have been selected over the period 2018-2020 with energies $>60$\,TeV, making them the best astrophysical candidates so far.
Furthermore, multi-messenger follow-up have been set, and alerts are planned to be sent soon.

\vspace{-0.2cm}
\paragraph{Sea experiments}
Two experiments have initiated the efforts towards the deployment of detectors in deep sea: {\bf NESTOR} ({\bf Neutrino Extended Submarine Telescope with Oceanographic Research Project}), and {\bf NEMO} ({\bf Neutrino Mediterranean Observatory}). 
NESTOR was located off the Greek coast at about $3800$\,m depth in the Ionian Sea, and NEMO was deployed close to Sicily at $100$\,km from Capo Passero.
Both experiments measured the atmospheric muon fluxes and paved the way for a larger scale detector.

{\bf ANTARES} ({\bf Astronomy with a Neutrino Telescope and Abyss Environmental Research}) was the first water detector with a size comparable to AMANDA.
It consists of $12$ strings anchored to the sea bed, and kept vertical thanks to buoys.
Strings are separated by about $60$\,m, each string is composed of $25$ storeys, spaced by $14.5$\,m, with the lowest at $100$\,m above the sea bed and the highest at about $460$\,m above.
Each storey is composed of $3$ 10-inch PMT.
The storeys are connected with electro-optical cables (21 optical fibers for digital communications).
Each string is divided into $5$ sectors, each containing $5$ storeys.
Storeys are controlled by a Local Control Module (LCM) which handles data communications between its sector and the shore station.
The signals from PMTs are digitized with a sub-nanosecond precision, thanks to an interplay between the clock of the LCM and the master clock at shore.
Time calibration is performed thanks to pulses between shore clock and LCM clocks, and with LED beacons that fires at the same time the digitizing system electrically and the PMTs optically.
Because the strings are immersed in sea current, their positions can vary in time, therefore a calibration of the positions is achieved thanks to compasses in each storey, tilt meters along the strings, and an acoustic triangulation system, composed of transmitters at the bottom of the strings, and hydrophones along the strings. An overall precision of a few cm on the relative positions of OMs is achieved.
The energy threshold is $20$\,GeV for tracks and $1$\,TeV for cascades~\cite{Guepin_2022}.
The angular resolution, estimated from Monte-Carlo simulations for muon tracks, is $\sim0.2\degree$ at $10$\,TeV, $0.7\degree$ at $1$\,TeV, and $1.8\degree$ at $100$\,GeV. Towards lower energies, kinematics effects between the neutrino and the muon limit the resolution.
For cascades, a median mismatch of $\sim 10\degree$ is relatively easily achieved, while with refined methods and cuts a resolution down to $\sim3\degree$ can be reached. 
Finally, the full configuration of ANTARES was completed in 2008, and it was switched off in 2022. 

The next generation of European large scale water Cherenkov detector is the {\bf cubic KiloMeter cherenkov NEutrino Telescope} ({\bf KM3NeT}). It is currently under deployment at the bottom of the Mediterranean Sea, at two main locations: offshore Toulon in France ($\sim2450$\,m), and Capo Passero in Sicily in Italy ($\sim3500$\,m). The complete detector consist of 3 building blocks, each of the building blocks are made of $115$ strings with $18$ DOMs each.
Offshore Sicily, the detector is called {\bf ARCA} ({\bf Astroparticle Research with Cosmics in the Abyss}) ; it focuses on the study of high-energy cosmic neutrinos in the similar energy range as IceCube. 
It will consist of 2 building blocks in a sparse layout, in order to reach a detector volume of a cubic kilometer. It will complement the field of view of IceCube by looking at the Southern Hemisphere, where the galactic center is visible. Furthermore, a better resolution on the angular reconstruction of the tracks is expected, compared to IceCube, because the scattering length is larger in water, compared to the ice.
The targeted energy threshold is $\sim 100$\,GeV for tracks and $\sim 1$\,TeV for cascades~\cite{Guepin_2022}. 
The envisioned resolution on the angular reconstruction is $0.1\degree$ at $1$\,PeV for muon tracks, and $\sim1.5\degree$ for cascades~\cite{Guepin_2022}. 
Offshore Toulon, the detector is called {\bf ORCA} ({\bf Oscillations Research with Cosmics in the Abyss}). It is composed of the third building block with a high density layout, in order to focus on low energy neutrinos (down to $\sim4$\,GeV) produced in the atmosphere for precise oscillations measurements.
Each DOM is made of a $43$\,cm pressure glass sphere, with $31$ PMTs of $7.5$\,cm each. This configuration has several advantages compared to a single large PMT: the photocathode area is more than three times larger than the one of a single $25$\,cm PMT.
The individual readout from each PMT allows for a good separation between different photoelectrons, hence it helps to filter the data. Finally, the configuration provides  information on the direction of emission. This concept was validated with in-situ prototypes.
For each pulse seen by a PMT, after passing some preset thresholds, the leading edge time and time over threshold are digitized and send as hit, instead of digitizing the whole waveform. 
Therefore, each hit is only 6 bytes of data. All hits are sent to shore, following the concept called "All data to shore".
The total rate of data from a single building block (of $64,170$ PMTs) is $\sim 25$\,Gb/s, sent through optical fibers to shore, using wavelength multiplexing to optimize the data transfer.
Once at shore, each event is filtered and discriminated against noise, resulting in a reduction factor of $10^5$, compared to the flux of data arriving. Each saved event contains all the data during that event, in order to have the maximum of information for off-line analysis, and are saved on disks.

KM3NeT is considered as the European counterpart to IceCube, focusing on the Southern Hemisphere. 
In principle, a complete ARCA detector could detect IceCube astrophysical neutrino flux at a $5\,\sigma$ level in $1$ year of operation. 
Thanks to its location, its best sensitivity is toward the Galactic center, where several $\gamma$-ray TeV sources are detected. In less than $4$ years, the predicted neutrino flux should be probed, and constraints set. Furthermore, the good angular resolution, as well as its field of view, should allow for multi-messenger followups.
Finally, ORCA could determine the neutrino mass hierarchy with a significance of $3\,\sigma$ in $3$ years of data taking.

\paragraph*{}\vspace{-0.2cm}
To complete the picture, let us mention P-ONE (Pacific Ocean Neutrino Experiment) that builds on a similar concept as KM3NeT and Baikal GVD but deployed in the Pacific Ocean. It will be able to increase the event statistics from the Southern Hemisphere and complement the field of view and followups from the current experiments (see~\cite{SnowMassHENu_2022} for more details). Last but not least, IceCube-Gen2 is a large and multi-instrument extension planned for IceCube, and is detailed in section~\ref{sec:UHE}.

\vspace{-0.2cm}
\subsection{A quiet neighborhood}
\vspace{-0.2cm}
In 2013 IceCube has detected for the first time a diffuse astrophysical neutrino flux, now with a significance larger than $7\,\sigma$: a breakthrough in the field. 
The origin of this diffuse flux still remains unclear.
Recently, three major discoveries, made by IceCube, have shed new lights on this enigma.
In 2018, a possible detection of a neutrino in coincidence with electromagnetic radiations (X-ray, $\gamma$-ray, and optical) was evidenced above $3\,\sigma$ for the first time. The neutrino was identified as coming from the direction of the blazar TXS0506+056, in an active state at that time. 
This result hinted that active-galactic-nuclei could be a source of high-energy neutrinos. 
It was confirmed, last year, by the discovery of a neutrino excess in the direction of the active-galactic-nuclei NGC1068.
This excess was observed over a time period of $3186$ days, and represents $79^{+22}_{-20}$ neutrinos above the atmospheric and cosmic neutrino backgrounds, leading to an significance of $4.2\,\sigma$. The neutrino flux associated to the source is $\Phi^{\rm 1 TeV}_{\nu_\mu + \bar{\nu}_\mu} = \qty(5.0 \pm 1.5_{\rm stat}) \times 10^{-11}$\,TeV$^{-1}$\,cm$^{-2}$\,s$^{-1}$. 
This flux can be converted to a total neutrino luminosity at the source, taking into account the flavor ratio, emission mechanism and distance to NGC1068.  The isotropic and redshift corrected neutrino luminosity is $L_\nu \qty(1.5\to15\,{\rm TeV}) = \qty(2.9\pm1.1_{\rm stat}) \times 10^{42}$\,erg\,s$^{-1}$. Interestingly, this neutrino flux is more than a factor $10$ higher than the equivalent $\gamma$-ray luminosity observed in the energy range of $100$\,MeV to $100$\,GeV (and higher than the limits above $200$\,GeV), and which is $L_\gamma = 1.6 \times 10^{41}$\,erg\,s$^{-1}$. This result suggests a dense environment around or within the source, which absorbs the $\gamma$ rays, and not the neutrinos.
The consequence of this discovery, added to the evidences from TX0506+056, is that active-galactic-nuclei could contribute significantly to the overall diffuse neutrino flux, considering that their individual contribution to this flux is around $1\%$ in their respective energy range. It has to be noted that the two sources aforementioned are intrinsically related to active-galactic-nuclei, but most likely emitted neutrino via different mechanisms: the former was a blazar in a flaring state, while the latter is a Seyfert Galaxy and a steady state neutrino emitter.

Finally, this year, the detection of neutrinos coming from the Galactic plane, has evidenced a new origin for a part of the observed diffuse neutrino flux.
Indeed, a diffuse neutrino emission from the galactic plane is expected from the interactions of cosmic rays inside the galaxy, and producing neutrinos. From their detection, it is possible to locate the interaction sites and infer the energetics at play.
IceCube's location in the Southern Hemisphere is not ideal to observe the Galactic Center, due to atmospheric muon tracks that pollute the expected signals from the muon tracks produced by astrophysical neutrinos. However, by using cascade events it is possible to greatly reduce the expected background.
Thanks to a novel hybrid method, which involves a complete parametrization of the detector response with a neural network, the number of retained events could be increased by a factor $20$ (from $1980$ to $59592$, from $500$\,GeV to several PeVs). Furthermore, the angular resolution could be improved by a factor $2$ at TeV energies.
Consequently, a diffuse neutrino emission was observed from the Galactic Plane with a significance of $4.71\,\sigma$ for the best model~\cite{IceCube_2023}.

In the last decade, IceCube has made significant discoveries from the Northern sky, supported by the constraints set by ANTARES from the Southern sky. Their successors, namely IceCube-Gen2 and KM3NeT, should deepen these results by significantly increasing the number of neutrinos detected.
Finally, no cosmogenic neutrinos, expected from the interaction of ultra-high-energy cosmic rays with cosmic photons, have been discovered yet. New detectors of multi cubic kilometers seem to be needed to reach the expected fluxes from this component of the ultra-high-energy astrophysical neutrinos.

\vspace{-0.2cm}
\section{Ultra-high-energy neutrinos: an uncharted territory} \label{sec:UHE}
\vspace{-0.2cm}
The neutrinos fluxes become extremely low towards ultra-high-energies, and requires effective volumes up to hundreds of gigatons, to achieve detection. As an illustration, a detector scale larger than $100$\,km$^3$ is needed to detect more than a handful of cosmogenic neutrinos, with a typical energy range of $100$\,PeV to $10$\,EeV.
In principle, such detectors can be designed with the same technique and strategy as the detectors of high-energy neutrinos. Technically, the monitoring of several cubic kilometers of ice and sea can be conducted with the standard detection technique of the Cherenkov tracks and cascades induced by neutrinos. However, the practical deployment of such detector become extremely cost-constraining. Therefore, many projects have investigated alternative methods which, in fact, are already used and matured by other fields of the astroparticle community.

\vspace{-0.2cm}
\subsection{Alternative detection methods}
\vspace{-0.2cm}
The detectors need to monitor natural targets as big as possible, in order to reach the low fluxes at ultra-high-energies.
The biggest accessible volumes on Earth are: the Earth itself, the atmosphere layers and the polar ice caps. 
On a side note, it is also possible to monitor the Moon from the Earth, and it has been investigated by a few experiments over the past decade. For these volumes, scaling-up the standard techniques starts to become cost-constraining (but doable in principle). 

Interestingly, new techniques become competitive with the standard Cherenkov tracking, such as the air-shower imaging (already well established by the cosmic-ray community), and radio detection (which has an excellent duty cycle).
These methods no longer rely on the detection of tracks produced by neutrino induced lepton, but instead focus on the detection of the particle cascade (shower) resulting from the decay of the aforementioned lepton. Ultra-high-energy particles have a higher chance to decay and induce a particle shower, no matter the detection medium used.
In the air, these showers, when produced by Earth-skimming neutrinos, propagate over 10s to 100s of kilometers and emit Cherenkov radiation, fluorescence light and electromagnetic radiation, all detectable on Earth, alongside with the particles themselves reaching the ground. In denser media, such as ice or water, the showers extend over smaller distances, of the order of a few meters for the longitudinal profile and of a few centimeters for the lateral one. Nevertheless, these showers also emit Cherenkov radiation, and even radio waves, except for liquid water environments.

\vspace{-0.2cm}
\paragraph{Cherenkov imaging}
Particles from an air-shower move at relativistic speeds, therefore, when propagating in the atmosphere, produce Cherenkov light along their path.
The imaging technique relies on the use of so-called Cherenkov telescopes, similar to the telescopes used in standard astronomy: made of a primary mirror recorded by pixelated cameras. The system aims at imaging the Cherenkov tracks, of nanoseconds scale, seen in the atmosphere by dark moonless nights. Therefore, this technique makes use of a vast natural medium, the atmosphere, to produce Cherenkov light. However, due to possible light contamination from the Sun, the Moon, and human activities, the duty cycle is rather limited. This technique has been successfully used on ground based telescopes, onboard flying stratospheric balloons, and is envisioned to be deployed in space.

\vspace{-0.2cm}
\paragraph{Fluorescence imaging}
Extensive air-showers produce fluorescence light when traveling through the atmosphere. The charged particles of the shower, mainly electrons and positrons, deposit energy within the molecules of the air, under the form of ionization and excitation~\cite{Risse_2003}. Most of the fluorescence of an air-shower results from the excitation of two electronic states from the nitrogen molecule.
Some of this excitation energy is then released as visible and U.V. light, where the emission peaks in the $300-400$\,nm band. 
The fluorescence light technique allows for the most direct measurement of the development of the longitudinal profile of the air-shower~\cite{Keilhauer_2005}. Therefore, it is very well suited for the reconstruction of the energy and the direction of the primary particle. However it is critical to correctly understand the local atmospheric conditions, and the modeling of the photon yield from fluorescence. The photon yield connects the detected photons to the energy deposited in the atmosphere by the shower, to reconstruct the energy of the primary particle. This quantity is highly dependent on the atmospheric conditions, where for instance, emitted light undergoes Rayleigh scattering with the atmosphere molecules, which absorbs a part of the photon energy.
Similarly to Cherenkov imaging, this technique has been successfully used on ground, onboard flying balloons, and is envisioned to be deployed in space.

\vspace{-0.2cm}
\paragraph{Radio detection}
The interaction of the particles from the shower with its environment also leads to the emission of radio waves~\cite{Schroder_2016}. The macroscopic features of this radio emission highly depends on the type of media where the cascade develops, since it will affect the typical size of the shower and the propagation of the radio emission. For instance, in-ice showers have dimensions on the order of tens to hundreds of centimeters, while air-showers reach tens to hundreds of kilometers. In particular, the particle front of the shower, also called "pancake", has a typical thickness of a few centimeters in ice (a few meters in air), and tens of centimeters of diameter in ice (tens of meters in air).
For air-showers, the radiation results from two main mechanisms, with an intensity peaking in the $ \rm MHz$ regime:
\begin{itemize}[noitemsep]
\item[1] The geomagnetic emission: it is due to the deflection of the lightest charged particles in the shower, i.e., positrons and electrons in opposite directions, due to the Lorentz force. This force induces a current varying in time as the particle content in the shower varies over time, leading to a radio signal polarized along the $-\mathbf{v} \times \mathbf{B}$ direction (with $\mathbf{B}$ being the direction of the magnetic field and $\mathbf{v}$ the direction of the shower). 
\item[2] The charge-excess or Askaryan emission: while the shower propagates, electrons from air-molecules (or water molecules in ice) are struck by high-energy shower particles and then travel along with the shower front. This combined with positron annihilation leads to a build up of a net negative charge in the shower front.
This negative charge excess can be up to $20-30\%$ and induces a dipole between the positively charged plasma behind the shower front and the electrons in the front, inducing a signal radially polarized in a plane perpendicular to the shower axis.
\end{itemize}
The geomagnetic emission is dominant in the air, where the charge-excess account for only $\sim 1-20\%$ (depending on the geomagnetic orientation) of the signal. But it is negligible in denser media, such as ice (or rock) where the charge-excess corresponds to the dominant emission mechanism. This can be explained by the extension of the shower and the density of surrounding medium.
At a specific angle, the radiations from the shower arrive simultaneously at the observer. The observed signal is therefore composed of a very brief and intense pulse in the time domain, and an extended emission in the frequency domain~\cite{Schroder_2016}. This geometrical time compression effect is called Cherenkov effect, and confines the emission in a cone with a typical aperture angle (called Cherenkov angle) of $1\degree$ for an air-shower and of roughly $40\degree$ to $60\degree$ for an ice-shower.
For extensive-air-showers, the coherence of the signal is maintained outside the Cherenkov cone, even though the radio pulses get broader due to the difference of optical paths. For the same reason, but in a more critical regime because of the larger refractive index, the coherence is lost for in-ice radio emissions outside the Cherenkov cone.
The radio detection of the emission from particle showers has been extensively studied over the past decades, and benefits in particular from the long-lasting experience of the radio telecommunication and radio astronomy fields (see e.g~\cite{Balanis_2012} for a complete review). Thanks to the great attenuation length of radio waves, both in air ($\sim 1000$\,km) and in ice ($\sim 1$\,km) this detection technique is competitive and even superior to optical techniques above $10$\,PeV.

\vspace{-0.2cm}
\subsection{Alternative detection strategies}
\vspace{-0.2cm}
The great variety of environments and topographies around the Earth has been used to investigate different concepts of detection strategies and techniques: from deep ice caps, up to space.

\vspace{-0.2cm}
\paragraph{Radio detection in the ice}
The ice provides a denser interaction medium than the air. 
Which, in principle, results in a larger effective volume for the radio detection technique in the ice compared to the in-air technique.
However, this is balanced by the shorter attenuation length of radio waves in the ice compared to the air. In addition, the emission is only coherent along the Cherenkov cone, and the variations of refractive index with depth leads to ray-bending and refraction effects. All these effects can reduce the effective volume for radio detectors in ice and complicate the reconstruction and the interpretation of the data.

The ice caps offer a gigantic interaction volume for neutrinos, which makes possible, in principle, the deployment of a detector with an effective volume of the order of several gigatons, in a relatively radio quiet environment.
This detection strategy has been investigated at the South Pole, by the pioneering experiments {\bf RICE} ({\bf Radio Ice Cherenkov Experiment}) 1995-2005, and {\bf AURA} ({\bf Antarctic Under-ice Radio Array}) 2003-2009~\cite{Kravchenko_2003,Landsman_2009}. They have demonstrated the feasibility of the technique for two concepts: subsurface and deep ice antenna arrays. The sub-surface arrays while being obviously more convenient for the deployment, faces more refraction and ray-bending effects than deep-ice ones, where the ice is colder and more stable.
Following these two concepts, the {\bf ARA} ({\bf Askaryan Radio Array}) and {\bf ARIANNA} ({\bf Antarctic Ross Ice Shelf Antenna Neutrino Array}) experiments~\cite{Allison_2016,Anker_2019} applied the same techniques at a larger scale.

ARA is a deep-ice antenna concept, operating since 2010 near the Amundsen-Scott station at the South Pole, and is the evolution of AURA.
Each station is situated at $\sim200$\,m depth, and made of $4$ strings composed of $16$ cylindrical antennas in interferometer mode, and operating in the $200-850$\,MHz band, for a volume of $20\times20\times20$\,m$^{3}$.
The energy threshold is about $\sim10$\,PeV, for an angular reconstruction of $\sim5\degree$~\cite{Guepin_2022}.
The $37$ stations currently deployed, in a hexagon with a $2$\,km spacing, lead to an effective volume of $200$\,km$^{3}$ at $1$\,EeV. This volume, while being the largest, monitored by in-ice radio antennas remains well below the required values since it would need a volume at least $4$ times larger in order to reach the ultra-high-energy fluxes predicted by cosmogenic models.

ARIANNA is a sub-surface antenna experiments, originally located in Moore's bay at $110$\,km from the Mc Murdo station, and is in operation since 2012. Its goal is to observe the ice layer of $\sim 570$\,m above the Ross sea.
The initial detector concept relies on a layout of $36\times36$ stations with a spacing of $1$\,km step. Each station is made of $8$ down-looking and $2$ upward looking antennas (for calibration and cosmic-ray veto), and operates between $100-1300$\,MHz.
The detector profits from the radio reflections at the interface between water and ice at the bottom of the ice shelf. It allows for detecting both direct and indirect signals, leading to an increase in the field of view and effective volume, by a factor of almost $2$.
However, since the antennas are deployed close to the upper layers of the ice (the firn) the attenuation length is only of about $400-500$\,m.
Consequently, the energy threshold is of order $30$\,PeV, and the angular resolution is between $2.9\degree-3.8\degree$~\cite{Guepin_2022}.
A hexagonal array of pilot stations has been deployed since 2012 and gradually completed with up to $7$ stations, with a few test stations deployed in 2018. In addition, two stations have also been deployed at the South Pole.
Yet, only a few dozen of stations were deployed in total, far from the $1000$ stations initially needed to reach the fluxes expected for cosmogenic neutrinos.

The two concepts depicted above have recently merged to combine resources and advantages in the {\bf Radio Neutrino Observatory in Greenland}~\cite{Aguilar_2021} ({\bf RNO-G}).
It relies on the developments of both ARA and ARIANNA, in order to reach the effective volume needed to detect ultra-high-energy neutrinos, while minimizing the required number of antenna stations.
The array is located at the Summit Station, profiting from the layer of $3$\,km of deep ice in central Greenland.
Each station is composed in total of $24$ antennas: a deep-ice log-periodic array ($150-600$\,MHz), a la ARA, designed with a phased trigger system, and a set of subsurface antennas ($100-1300$\,MHz), a la ARIANNA, oriented in order to be able to fully measure the polarization of any signal induced by a shower. The combination of the two techniques should increase the reconstruction capabilities of the station.
The deep-ice array is made of $3$ strings plunging into the ice sheet, down to $100$\,m and monitors a volume of ice of roughly $1$\,km$^3$. 
The expected energy threshold is $50$\,PeV, and the envisioned angular resolution is $\sim 2\degree \cross 10\degree$~\cite{Guepin_2022}.
The final configuration should be made of $35$ independent stations separated by $1.5$\,km. 
The design of RNO-G will serve as a reference to build the IceCube-Gen2 radio detector (see the last paragraph of this section).

\vspace{-0.2cm}
\paragraph{Radio detection from the stratosphere}
The radio signals induced by particle cascades in the ice can be refracted up to the surface, and propagate over long distances thanks to the large attenuation length of radio waves in the air.
Following this idea, it is possible to benefit from a high altitude standing point to monitor a huge volume of ice. At stratospheric altitudes (on average $40$\,km), it is possible to scan up to $650$\,km away, providing an equivalent detector volume of a million gigaton of ice, at a cost of a threshold on the neutrino energy in the EeV range.

This strategy was followed by the {\bf ANITA} missions ({\bf ANtarctic Impulsive Transient Antenna}, see e.g.,~\cite{,Gorham_2019} and references therein). It is a series of NASA missions, which consisted in a few flights of stratospheric balloons above the ice cap in Antarctica, profiting from the wind vortex at the South Pole.
The detector onboard the payload was designed to detect both in-ice and in-air showers, by measuring the signals from geomagnetic (in air) and Askaryan (in ice) radio emissions. Its location in the atmosphere allows for detecting both down-going and up-going trajectories, and both direct and reflected signals. A total of 5 missions (2006, 2008, 2009, 2014 and 2016) were launched with successive improvements on the design.
ANITA IV, lasted for $\sim30$\,days, with the following payload detector:
the radio instrument was made of $48$ quad ridged horn antennas, dual polarized, and operating in the $180-1200$\,MHz band. The array layout was designed on $3$ cylindrical layers, covering the complete azimuth range in 16 sectors. The RF signal chain allowed for a "threshold riding" trigger system constantly adjusting to the background with an event rate up to $50$\,Hz. 
The energy threshold was $0.1$\,EeV, and $2.8\degree$ for the angular resolution~\cite{Guepin_2022}.
The offline analysis made use of the interferometric technique between antennas, to improve the signal-to-noise-ratio and the angular resolution down to $0.1-0.2\degree$ (Askaryan) / $1\degree$ geomagnetic.

Previous ANITA flights have observed anomalous up-going events~\cite{Gorham_2021} with very steep trajectories. These events do not show any phase inversion, unlike other cosmic rays observed below the horizon (due to the ice reflection), and still remain to be explained. 
During the last ANITA (IV) flight, $27$ cosmic-ray events were clearly identified, among which $23$ have the expected polarity from their geometry. However, $4$ near-horizon events do not present any polarity inversion. Therefore, these events are inconsistent with cosmic-ray signals reflected off the ice, even though, they are identified as coming out of the ice sheet. Two of these events, present only a $1-2\,\sigma$ significance on their arrival direction, and thus could, in fact, originate from above the horizon. In which case, they would be compatible with cosmic rays. However, constraints on the propagation and coherence of the radio signal make this hypothesis unlikely. A detailed simulation study over the complete set of detected cosmic rays have shown a significance of $3.3\pm0.5\,\sigma$ regarding the detected anomalous events. The confidence level does not allow for a firm conclusion, however it suggests a new class of cosmic-ray-like events with Earth-skimming trajectories.
Another possibility, is that these events were produced by the decay in the atmosphere of a tau-lepton induced by an Earth-skimming tau neutrino. If that is the case, current limits on the diffuse flux of tau neutrinos would rather suggest a point source origin.

In addition, to these unprecedented discoveries, the ANITA missions have set the most stringent constraints on astrophysical neutrinos at GZK energies: $E^{-2}\times 1.3 \times 10^{-7}$\,GeV.cm$^{-2}$.s$^{-1}$.sr$^{-1}$ at $90\%$C.L. for an $E^{-2}$ spectrum in the range of $E_\nu=10^{18}-10^{23.5}$\,eV.

The future mission of this kind is called {\bf PUEO}~\cite{Abarr_2021} ({\bf Payload for Ultrahigh Energy Observations}). It plans to increase by almost two orders of magnitude the combined sensitivity of all previous ANITA missions (from ANITA I to ANITA IV), while keeping the same constraints in terms of detector size (because of the balloon requirements). To tackle this challenge, PUEO will improve the ANITA detector design with recent developments made in hardware and firmware, and proceed to the break-down of the instrument in two sub-instruments:
\begin{itemize}[noitemsep]
\item the main instrument is composed of $108$ quad-ridged horn antennas (twice more than ANITA IV), in the $300-1200$\,MHz band. In addition, $12$ antennas canted in the direction of the ground, and dedicated to disentangle the very steep trajectories. Furthermore, it will run on a phased trigger system, allowing for a drastic reduction of the trigger threshold.
\item the low frequency instrument is composed of $8$ sinuous antennas in the $50-300$\,MHz band. This instrument is dedicated to the detection of the radio emissions from extensive-air-showers, induced by Earth-skimming tau neutrino. The effective area is twice larger than the main instrument, for this channel. These performances are achieved thanks to the large collecting aperture of each antenna (about $1.9$\,m in diameter). As well as the frequency band, for which, the radio beam induced by the air-shower is larger by almost a factor $2$ compared to the frequency band of the main instrument.
\end{itemize}
Thanks to the split of the frequency band, the number of channels is doubled for the same detector size, and the radio noise (more intense in the low frequency band) is restricted to only one instrument.
The expected energy threshold will be similar to ANITA IV, and the online angular reconstruction will be improved thanks to the phasing of the antennas.
Finally, PUEO is planning to fly during the austral summer of 2025-2026 from Mc Murdo, for a nominal flight time of $30$ days.

\vspace{-0.2cm}
\paragraph{Imaging from the air and space}
Stratospheric balloons have also been investigated to monitor the atmosphere itself, thanks to the Cherenkov and fluorescence imaging. The large volume of atmosphere compensates for the more constrained duty cycle imposed by these light detection techniques, which require very dark environments. 

The {\bf Extreme Universe Space Observatory} ({\bf EUSO}) is a succession of near-space and space based missions. Their goals were to validate the detection strategy of POEMMA (see below), which relies on the detection of fluorescence and Cherenkov emissions from ultra-high-energy particles~\cite{Olinto_2021}. These detection techniques have been intensively used by cosmic-ray experiments such as the {\bf Pierre Auger Observatory}, and the {\bf Telescope Array}, which have also investigated the potential for neutrino detection through the direct measurements of the particles content from neutrinos induced air-showers (see e.g.,~\cite{SnowMassHENu_2022}).
The last mission of this type called {\bf EUSO-SPB2} (for {\bf EUSO aboard a Super Pressure Balloon 2}), is the follow-up mission of {\bf EUSO-Balloon} (2014) and {\bf EUSO-SPB1} (2017)~\cite{Eser_2021}. The mission is composed of two telescopes:
the fluorescence telescope, pointing downward and measuring micro-second scales fluorescence lights from ultra-high-energy cosmic-ray tracks ; and the Cherenkov telescope, pointing towards the limb and measuring nanoseconds scales Cherenkov emissions produced by Earth-skimming neutrinos.
It was launched from Wanaka in New-Zealand during the spring 2023.
The goals of this mission were: to quantify the air glow background from the night-sky near the Earth's limb for a future space mission ; and to measure 100s of direct cosmic rays per hour, and to test the reconstruction procedures. Finally, a program of target of opportunity was also planned for follow-ups of transient events of high-energies. Unfortunately, due to a hole in the balloon of the NASA, the payload could not complete its mission, and only lasted for a few hours. Nevertheless the technology could still be tested and validated up to some extent. Furthermore the NASA offered a new flight in compensation and the collaboration is already developing a new payload for a future mission.

The {\bf Probe of Extreme Multi-messenger Astrophysics} ({\bf POEMMA}) is a NASA astrophysics probe-class space based mission, and a potential candidate for a future NASA probe announcement of opportunity~\cite{Olinto_2021}. 
Its goal is to measure ultra-high-energy cosmic rays and cosmic neutrinos, by using the wide field of view in combination with the Earth and its atmosphere as neutrino targets. To do that, it aims at detecting the optical signals from extensive-air-showers resulting from neutrino interaction.
The design of the mission relies on two identical spacecrafts flying in a loose formation with a separation of $300$\,km at an altitude of $525$\,km, with an orbit inclined by $28.5\degree$. Each spacecraft will be composed of a Schmidt telescope with an optical collecting area of $6$\,m$^2$ and a field of view of $45\degree$. The focal plane of each telescope will be divided into 2 sections: one, optimized for a fluorescence camera and recording $80\%$ of the mirror ; and the other, optimized for a Cherenkov camera (recording $20\%$ of the mirror). The fluorescence camera will focus on the fluorescence light emitted by ultra-high-energy cosmic rays, and the Cherenkov camera will be focusing on the Cherenkov light emitted by Earth-skimming tau neutrinos. For the latter, the telescopes have to be sufficiently tilted in order to watch the Earth's limb.
The planned energy threshold and angular resolution are $10$\,PeV and $0.4\degree$ for the Cherenkov channel, while for the fluorescence channel they should be higher with $10$\,EeV and $1\degree$~\cite{Guepin_2022}.
The separation between the spacecrafts can be reduced to $25$\,km to observe the light from air-showers going upward, hence reducing the energy threshold for the detection of the neutrinos. The stereo observation mode allows for a more precise monitoring of $10^{4}$ gigaton of atmosphere. The telescopes can repoint within $8$ minutes, allowing for efficient searches of follow-up events across the sky. Finally, the orbital period is about $\sim 95$\,minutes, providing a full sky coverage for sources of ultra-high-energy cosmic rays and neutrinos.

\vspace{-0.2cm}
\paragraph{Radio detection from the ground}
Ground based radio-detectors are also envisioned to detect ultra-high-energy neutrinos. The detection strategy relies on the possibility to monitor gigantic volumes of atmosphere to catch the air-showers induced by the tau decays. In order to reach this goal, it must be demonstrated that sparse and extended arrays can be deployed and operated autonomously.
Several radio experiments have demonstrated the feasibility of air-shower detection (induced by cosmic rays): {\bf CODALEMA}, {\bf LOPES}, {\bf AERA}, and {\bf LOFAR}~\cite{Chiche_2022}. However, autonomous radio detection was investigated by a few of them only, such as {\bf AERA} and {\bf TREND} (the {\bf Tianshan Radio-array for Neutrino Detection}~\cite{Charrier_2018}) 2009-2013. TREND took advantage of the radio quiet environment and the mountainous topography of a remote valley in the Tianshan mountains in Western China, to pave the way of the autonomous radio detection of air showers.

This achievement has opened the way to a large scale radio array called {\bf GRAND}~\cite{GRAND_2020} ({\bf Giant Radio array for Neutrino Detection}).
GRAND is a planned large-scale radio experiment dedicated to the detection of ultra-high-energy messengers, of energies above $50$\,PeV, with a main focus on ultra-high-energy neutrinos. It will consist of a radio array of $\sim 200\,000$ antennas over $200\,000$\,km$^{2}$ deployed in several mountainous regions around the world. The focus is put on very inclined air-showers to detect tau-neutrinos with Earth-skimming trajectories that go through a dense medium as a mountain or the Earth surface for up-going trajectories. 
The deployment of the GRAND experiment is expected to be staged, i.e., divided in 3 main steps, GRANDproto300, GRAND10k and GRAND200k. GRANDProto300 (GP300) is the deployment of the first 300 antennas over $\sim 200$\,km$^{2}$ in China, to detect cosmic rays and hopefully gamma rays in the $10^{16.5}-10^{18}$\,eV energy range. It will serve as a testbench for the GRAND experiment, validating the detection and the reconstruction feasibility of highly inclined showers ($\theta>80\degree$), by probing autonomous radio-detection on large-scale arrays, and angular reconstruction below $0.1\degree$. 
GRAND10k is expected for 2028, it will consist of two sites: GRAND North (Gobi Desert, China) and GRAND South (Argentina), with 5-10k antennas each, to work on issues related to large-scale arrays, and to detect the first ultra-high-energy neutrinos for optimistic fluxes. Two sites are ideal for a full sky coverage, and test various types of environments and related technical issues. 
Finally, GRAND200k will consist of 20 sub-arrays of 10 000 antennas all around the world, to reach the sensitivity necessary to ensure the detection of the ultra-high-energy neutrino fluxes.
At the moment, three prototypes are deployed: 
\begin{itemize}[noitemsep]
\item GP300 (300 antennas) in the Gobi Desert in China,
\item GRAND@Auger (10 antennas) at the Pierre Auger Observatory for cross-calibration,
\item GRAND@Nan\c{c}ay (4 antennas) at the Nan\c{c}ay Radio Observatory in France as a testbench.
\end{itemize}
With the prototypes, the collaboration will validate the autonomous radio detection technique, calibrate its antennas and develop efficient reconstruction methods for very inclined air-showers.

An alternative to the concept of sparse radio arrays is followed by {\bf BEACON}~\cite{Southall_2022} ({\bf Beam forming Elevated Array for COsmic Neutrinos}).
This project plans to use the radio interferometry technique in the $30-80$\,MHz range, to detect events induced by Earth-skimming tau neutrino. The concept aims at deploying antenna stations atop of high elevation mountains, in order to increase the field of view towards the ground. This strategy increases the collecting area of radio signals from events emerging below the horizon and propagating in upward trajectories.
Consequently, the expected energy threshold is $30$\,PeV, and the angular resolution should lay between $0.3\degree-1\degree$~\cite{Guepin_2022}.
BEACON is therefore building on two key elements: first the radio-interferometer technique, extensively used in radio-astronomy for observations with a high sensitivity, and second, its topography site, which provide a large field of view. Preliminary simulation studies have shown that, in principle, BEACON could reach the fluxes of ultra-high-energy neutrinos with a thousand antenna stations.
Currently, a prototype station, made of $8$ antennas, is deployed at the Barcroft Station in the White Mountains of California. The prototype is used as a test-bench for calibration and data analysis on cosmic-ray observations. From this prototype, a phase of gradual upscaling should follow and consists in the deployment of a thousand stations.

\vspace{-0.2cm}
\paragraph{Hybrid detector} \label{sec:hybrid}
In principle, nothing prevents from combining different detection methods within the same detector. This strategy is followed by {\bf IceCube-Gen2}~\cite{Clark_2021}.
The second generation of the IceCube Neutrino Observatory, will target neutrinos in the TeV to EeV energy range. In order to achieve that, it will rely on a design made of three subcomponents: an optical detector focusing on high-energy neutrinos, a large and sparse radio array targeting ultra-high-energy neutrinos, and a hybrid surface detector dedicated to the detection and the veto of cosmic-ray induced extensive-air-showers.
The optical component (Gen2-Optical) will consist of a detection volume of $8$\,km$^3$, including the existing optical array. It will be made of $120$ strings with $80$ DOMs per string (totaling $8160$ DOMs instead of $5160$ presently). Strings will be deployed between $1344-2689$\,m below the surface. The string spacing will be $240$\,m (instead of $125$\,m in present IceCube), in order to significantly increase the volume of the instrument, while maintaining an energy threshold ($\sim5$\,TeV) and reconstruction performances ($\sim0.3\degree-10\degree$) competitive~\cite{Guepin_2022}. Strings will be deployed in a sunflower pattern to improve the azimuthal homogeneity. Finally, DOMs will have an improved photon collection about three time larger than the current one, thanks to a multi-PMT design, inspired from other experiments.
The sparse radio detector (Gen2-radio) will extend the energy reach, up to the EeV regime, and will be located next to the optical Cherenkov detector. At the South Pole, the ice near the surface has an attenuation length of $2$\,km. The radio array will cover a surface of about $500$\,km$^2$. It will be composed of two types of radio detector: hybrid stations a la RNO-G at $\sim150$\,m below the surface, and antenna arrays a la ARIANNA, close to the surface with one dipole at $\sim 15$\,m below the surface. The expected energy threshold for the radio detector is $\sim 10$\,PeV for an angular resolution comparable to RNO-G~\cite{Guepin_2022}.

\paragraph*{}\vspace{-0.2cm}
Several other projects could not be detailed here and illustrate the great diversity of concepts in the field. They combine various detection strategy such as the interferometry for in-ice or lunar radio emission (e.g., {\bf TAROGE-M}, and {\bf SKA}), detection through radar echoes ({\bf RET}), arrays of Cherenkov tank deployed in deep valleys ({\bf TAMBO}), or ground based telescopes, using Cherenkov imaging ({\bf TRINITY}). The descriptions can be found in~\cite{SnowMassHENu_2022}.

\vspace{-0.2cm}
\section{Summary}
\begin{figure}[!ht]
\centering 
\includegraphics[width=0.9\columnwidth]{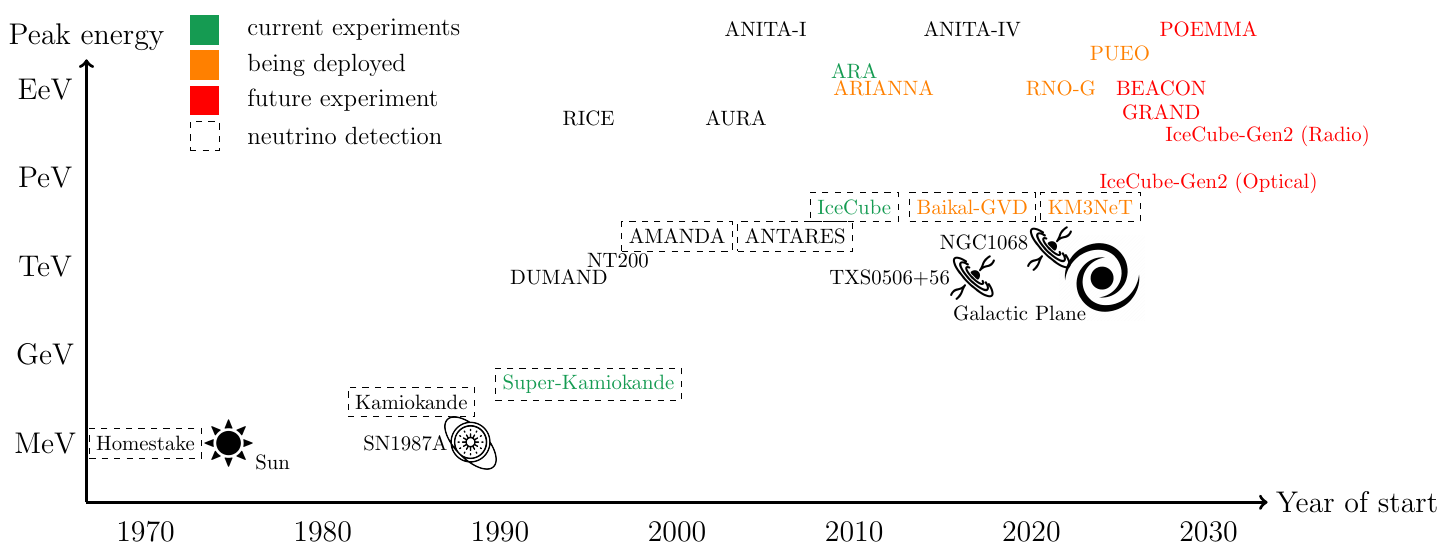}
\caption{Timeline of the detectors, and the few known neutrino sources, discussed in this review, as a function of their peak energy. Only a few of the many experiments designed have succeeded to detect the high-energy neutrino fluxes. A handful of sources have been identified, and the ultra-high-energy realm, remains untouched. This illustrates where the efforts for the next decades might focus.\vspace{-0.3cm}}\label{fig:summary}
\end{figure}

\vspace{-0.2cm}
Over the last decades, many technical and technological developments have been accomplished in order to detect the first astrophysical neutrinos.
One of the greatest challenges in such endeavor, beside the low interaction probability of the neutrino, is the large range of expected fluxes and energies. Consequently, the detector strategy and technique are tailored to each targeted energy range. As seen in this review, and illustrated in Figure~\ref{fig:summary}, these strategies and techniques vary greatly: from dense and compact underground detectors up to gigantic radio arrays on-ground and below ice, without forgetting near-space and space born instrumented payloads.
Thanks to these efforts, the neutrino sky starts to be visible to us. In particular, underground detectors have been the firsts to detect a Galactic supernova, and to confirm in-situ the chains of nuclear reactions taking place in the core of the Sun. However, their limited size and compactness, which grant us the access to these "low energy" neutrinos, cannot achieve the detection of neutrinos at higher energies. These neutrinos, in the high-energy regime, are expected from the interactions of accelerated particles, hence they hold some answers on the acceleration mechanisms within the sources. The neutrino sky, in this energy range, has been accessible thanks to the scaling of the underground detectors up to volumes of cubic kilometer scales, using natural targets such as ice, sea, and lakes. 
For these detectors, the origins of the neutrino sky remains largely unknown. Only a few sources have been identified yet, and many efforts have still to be conducted in order to reveal the origins of what is seen in the current data. Therefore, great perspectives are expected from the new detectors that are currently built, as they will increase the statistics of events and open new regions of the sky.
At the most extreme energies, no neutrinos have been detected yet. The uncertain predictions for these drastically low fluxes is a challenge for the design of the experiments. Consequently, a great diversity of detector concepts is being tested and matured. They push back the technical limits of astroparticle detections, and have renewed the detection strategies. Beside the expected and exciting detection of the first ultra-high-energy neutrinos, these experiments will also provide many technological advances and novel analysis methods.
Finally, a promising avenue for neutrino astronomy is the interplay with the multi-messenger astronomy. In addition to be a crucial element for multi-messenger astrophysics, neutrino detectors can greatly benefit from alerts and coincident detection, as illustrated with the case of TXS0506+56. Since in the high-energy and ultra-high-energy regime, transient sources are expected to contribute significantly, optimal synergies are required between the various neutrino detectors and the electromagnetic telescopes. The curious reader is strongly encouraged to look at the tables 1 and 2 from~\cite{Guepin_2022}, which review exhaustively these potential synergies. 
The neutrino sky just started to be revealed to us, and already high-energy astrophysics and particle physics has been marked by its imprint. In less than 30 years, neutrino experiments have evolved from a detector state to complete telescopes, and started the new field of neutrino astronomy.

\vspace{-0.2cm}
\subsection*{Acknowledgement}
\vspace{-0.2cm}
I wish to thank my closest scientific colleagues and friends for their careful readings, suggestions, and discussions: Claire Gu\'epin, Kumiko Kotera, and Olivier Martineau-Huynh. I also thank my colleagues Richard Dallier and Lilian Martin for discussion on the content of this proceeding.
\vspace{-0.2cm}
\begin{spacing}{.1}
{
\small
\bibliographystyle{JHEP}
\bibliography{bibliography}
}
\end{spacing}
\end{document}